\documentclass[twocolumn,showpacs,preprintnumbers,amsmath,amssymb,superscriptaddress]{revtex4}

\usepackage{graphicx}
\usepackage{dcolumn}
\usepackage{bm}

\begin{document}

\preprint{APS/123-QED}

\title{$^{13}$C NMR study of the magnetic properties of the quasi-one-dimensional conductor, (TMTTF)$_2$SbF$_6$}

\author{F. Iwase\footnote{Present address: Department of Physics, Okayama University, Okayama 700-8530, Japan}}
\affiliation{Institute for Molecular Science, Myodaiji, Okazaki, 444-8585, Japan}
\author{K. Sugiura}
\affiliation{The Graduate University for Advanced Studies, Myodaiji, Okazaki, 444-8585, Japan}
\author{K. Furukawa}
\affiliation{Institute for Molecular Science, Myodaiji, Okazaki, 444-8585, Japan}
\affiliation{The Graduate University for Advanced Studies, Myodaiji, Okazaki, 444-8585, Japan}
\author{T. Nakamura}
\affiliation{Institute for Molecular Science, Myodaiji, Okazaki, 444-8585, Japan}
\affiliation{The Graduate University for Advanced Studies, Myodaiji, Okazaki, 444-8585, Japan}

\date{\today}

\begin{abstract}
Magnetic properties in the quasi-one-dimensional organic salt (TMTTF)$_2$SbF$_6$ are investigated by $^{13}$C NMR under pressures. 
Antiferromagnetic phase transition at ambient pressure (AFI) is confirmed. 
Charge-ordering is suppressed by pressure and is not observed under 8 kbar. 
For $5 < P < 20$ kbar, a sharp spectrum and the rapid decrease of the spin-lattice relaxation rate $1/T_1$ were observed below about 4 K, attributed to a spin-gap transition.
Above 20 kbar, extremely broadened spectrum and critical increase of $1/T_1$ were observed. 
This indicates that the system enters into another antiferromagnetic phase (AFII) under pressure. 
The slope of the antiferromagnetic phase transition temperature $T_{\rm AFII}$, $dT_{\rm AFII}/dP$, is positive, while $T_{\rm AFI}$ decreases with pressure. 
The magnetic moment is weakly incommensurate with the lattice at 30 kbar.  
\end{abstract}
\pacs{71.20.Rv, 71.30.+h, 71.45.Lr, 76.60.-k}
\maketitle

\section{Introduction}
Quasi-one-dimensional (1D) organic conductors (TMT$C$F)$_2X$ ($C$=S, Se) are of interest for various electronic phases, such as charge ordering (CO), spin-Peierls (sP), antiferromagnetic (AF), incommensurate spin-density-wave (IC-SDW) and superconductivity (SC) by substituting anion $X$ and pressure \cite{Ishiguro,JeromeScience1991,JeromeCR2004}. 
Many body interactions including correlated electrons and electron-phonons are the origin of the variety of the phases.
A hallmark of the extensive experimental researches is a generalized phase diagram \cite{JeromeScience1991,ItoiJPSJ2008}. 
At high temperatures TMTTF salts are metallic state with 1/4-filled band, constructed from a TMTTF molecular $\pi$ orbital.
A metal-insulator crossover takes place by lowering temperature below about 200 K \cite{LaversanneJPL1984} due to Umklapp scattering \cite{EmeryPRL1982}.
In the insulating state, an anomaly in thermopower and resistivity was discussed in terms of {\it structureless transition} \cite{CoulonPRB1985}, which is now understood as the CO transition (or 4 $k_{\rm F}$ charge density wave, where $k_{\rm F}$ is the Fermi wave vector). 
Observations of an anomaly in dielectric permittivity \cite{NadEP1998,NadJPF1999,NadPRB2000} and line splitting of NMR spectrum \cite{ChowPRL2000} are the strong evidences of the CO transition. 
Analysis of the anisotropy of the ESR line width $\Delta H_{\rm pp}$ indicates that the CO pattern along the stacking axes is --O--o--O--o (where gOh denotes a charge-rich site and goh denotes a charge-poor site) \cite{NakamuraJPSJ2003}. 
The CO transition temperature, $T_{\rm CO}$, strongly depends on the size of anions, and also depends on the degree of the dimerization of TMTTF molecules or the strength of the electronic correlations \cite{NogamiJPF2005}. 
$T_{\rm CO}$ of PF$_6$, AsF$_6$, and SbF$_6$ are  65 K, 100 K, and 157 K, respectively. 
The driving force of the CO is regarded as the energy balance of the on-site Coulomb repulsion $U$, next-neighbor Coulomb repulsion $V$, and transfer integral $t$ \cite{HirschPRL1983, HirschPRB1983, SeoJPSJ1997}. 
On the other hand, it was pointed out that the effect of the Coulomb interaction is not enough to produce the CO state in the real material \cite{KuwabaraJPSJ2003}. 
Many theoretical studies have discussed that the CO state can be stabilized by the shift of the anions \cite{MonceauPRL2001, BrazovskiiSM2003, NadJPF2002, NadJPSJ2006, RieraPRB2000, RieraPRB2001}.
The displacement of anions is experimentally indicated by the structural investigation \cite{SouzaPRL2008}. 
However, the effect of the anions is secondary for the CO transition when we consider the behavior of an anion ordering (AO). 
ReO$_4$ salt shows the AO and CO separately at $T_{\rm AO}$=158 K and at $T_{\rm CO}$=225 K, respectively \cite{NakamuraJPSJ2006, NadJPF2002}. 
The deutration to TMTTF molecules increases $T_{\rm CO}$ whereas $T_{\rm AO}$ remains unchanged \cite{FurukawaJPSJ2005,NadJPCM2005}. 
This implies that the correlation between the CO and the anion is weak.
Applying pressure decreases the $T_{\rm CO}$ \cite{NadJPSJ2006,YuPRB2004,ZamborszkyPRB2002} due to the increase of $t$.

At the lower pressure portion of the generalized phase diagram presented in ref. \cite{JeromeScience1991}, there is the sP (or SG) phase \cite{FujiyamaJPSJ2006,DummJPF2004, LeylekianPRB2004,LeylekianPB2009, SouzaPB2009, FurukawaPRB2011}.    
(TMTTF)$_2$PF$_6$ shows the CO transition above the temperature of the sP transition.
In the sP phase, the CO state has been detected by the $^{13}$C NMR under the high magnetic field \cite{BrownPRL1998, BrownSM1999, ZamborszkyPRB2002}. 
Many theoretical studies proposed the possibility of the coexistence of the CO and sP \cite{KuwabaraJPSJ2003, RieraPRB2000, RieraPRB2001, ClayPRB2007}.
However, the amplitude of the charge separation considerably decreases in the sP phase \cite{FujiyamaJPSJ2006, NakamuraJPSJ2007, ClayPRB2007}.

\begin{figure}[b]
\includegraphics[scale=0.5]{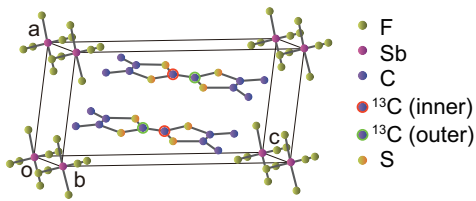}
\caption{Crystal structure of (TMTTF)$_2$SbF$_6$. Two carbons in a TMTTF molecule are $^{13}$C enriched. \label{fig:CrystalStructure}}
\end{figure}

The title complex (TMTTF)$_2$SbF$_6$ is expected to be located in the lower portion of the phase diagram based on its large lattice parameter \cite{ItoiJPSJ2008, YuPRB2004}. 
Therefore this complex is suitable for the study of the whole range of the phase diagram.
The crystal structure is shown in Fig. \ref{fig:CrystalStructure}. 
The unit cell has two TMTTF molecules including an inversion center ($P\bar{1}$).
The SC phase was observed in wide pressure range under extreme high pressures of 5.4 $<P\leq$ 9 GPa \cite{ItoiJPSJ2008}.
$T_{\rm CO}$ at ambient pressure is $\sim$157 K evidenced by dielectric susceptibility \cite{JavadiPRB1988, MonceauPRL2001,NadJPSJ2006}, ESR \cite{NakamuraJPSJ2003} and NMR \cite{YuPRB2004, YuJPF2004}. 
The ratio of the charge separation estimated by $^{13}$C NMR is about 3 : 1 \cite{YuJPF2004}.
The ground state at ambient pressure of (TMTTF)$_2$SbF$_6$ is an AF ($T_{\rm AF}\sim$7.5 K) \cite{MaaroufiMCLC1985, CoulonMCLC1985}. 
The striking observation of the AF phase rather than the sP phase at ambient pressure has cast doubt on the assumption that the dimensionality of the system increases with increasing pressure \cite{ItoiJPSJ2008, FurukawaPRB2011}.
The sP phase transition can be explained by the one-dimensional Heisenberg dimer model \cite{JacobsPRB1976}.
Therefore the forming of the sP phase by applying hydrostatic pressure as breaking the three-dimensional AF order is not easy to be understood, while the existence of the sP phase has been pointed out by two-dimensional $^{13}$C NMR \cite{YuPRB2004, YuJPF2004}.
The comprehensive studies of the magnetic properties of the electronic ground states and the detailed electronic interactions are required.
Second AF phase in (TMTTF)$_2$SbF$_6$ (AFII phase), which is expected at higher than the sP phase, have not been reported.

In this paper, the variation in the magnetic ground states of (TMTTF)$_2$SbF$_6$ are investigated by using standard 1D $^{13}$C NMR applying physical pressure. 
The CO is clearly observed by the NMR line splitting. 
$T_{\rm CO}$ decreases with increasing pressure.
At about 8 kbar the CO is not detected.
For 5 $<P<$ 20 kbar, a spin-gap state is observed in the reduction of the spin-lattice relaxation rate $1/T_1$ below about 4 K. 
Above 20 kbar, the ground state enters into another AF phase (AFII).
The broadened spectrum suggest that the magnetic moment in the AFII phase is weakly incommensurate with the lattice. 

\section{Experimental}
Single-crystal samples of TMTTF molecules with $^{13}$C at double-bonded carbons (see Fig. \ref{fig:CrystalStructure}) were prepared for NMR measurements using a standard electro-crystallization method \cite{FerraisTL1973, MasTL1977, BechgaardSSC1980, NakamuraJPSJ2003,NakamuraJPSJ2006}. 
A hybrid NiCrAl--BeCu clamp-type pressure cell was used to apply pressure. 
Daphne 7373 and Daphne 7474 were used as the pressure media for hydrostatic pressure in the pressure ranges of $P<$ 20 kbar and $P>$ 20 kbar, respectively. 
The pressures given in this report were measured at room temperature. 
The reduction in the internal pressure at low temperatures is less than 2 kbar for Daphne 7373 and 2.7 kbar for Daphne 7474 \cite{MurataRSI2008}. 
We developed a NMR probe which can rotate the sample around the molecular stacking $a$-axis with the pressure cell. 
An external field of 8 T ($\sim$86 MHz) was applied with at the magic angle such that additional line splitting originating from $^{13}$C=$^{13}$C dipolar coupling (Pake doublet) vanished. 
The NMR spectrum was obtained by taking the fast Fourier transform of the signal obtained after applying a $(\pi/2)$--$(\pi)$ spin-echo pulse sequence. 
The typical ($\pi/2$) pulse width is 2.8 $\mu$s. 
The spin-lattice relaxation rate $1/T_1$ was measured by the saturation recovery method.

\begin{figure}[b]
\includegraphics[scale=0.48]{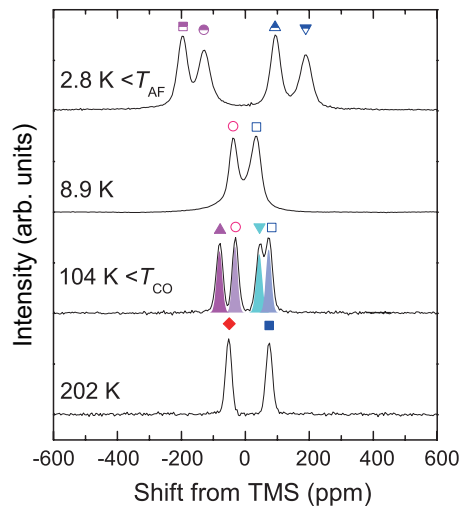}
\caption{Temperature dependence of the $^{13}$C NMR spectrum of (TMTTF)$_2$SbF$_6$ at ambient pressure. \label{fig:SPCAmbient}}
\end{figure}

\begin{figure}[b]
\includegraphics[scale=0.48]{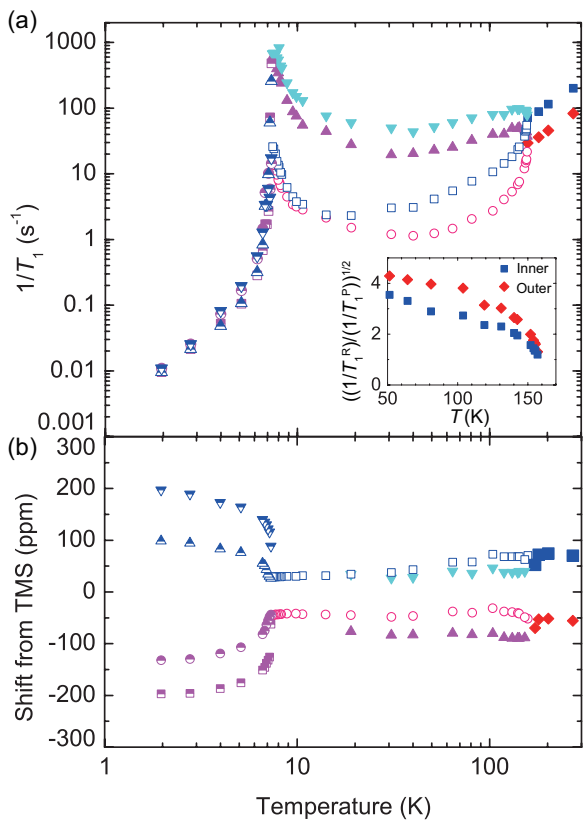}
\caption{(a) Temperature dependence of $1/T_1$. 
(b) Temperature dependence of the peak positions in the NMR spectrum. 
The NMR shift is measured from the tetramethylsilane (TMS) signal.
The symbols are the same as those in the spectrum in Fig. \ref{fig:SPCAmbient}.  \label{fig:Freq+T1Ambient}}
\end{figure}

\section{Results and discussion}
\subsection{Charge ordering and antiferromagnetic phase transition at ambient pressure}
Figure \ref{fig:SPCAmbient} shows the temperature dependence of the $^{13}$C NMR spectrum at ambient pressure.
The spectrum consists of two narrow lines at high temperatures. 
NMR shift, $K_{\rm obs}$, is expressed as $K_{\rm obs}=K_{\rm s}+K_{\rm orb}$, where $K_{\rm s}$ is the Knight shift, and $K_{\rm orb}$ is the chemical shift. 
$K_{\rm s}$ is related to the spin susceptibility $\chi_{\rm s}$ by the hyperfine coupling constant $A$ as $K_{\rm s}=A\chi_{\rm s}$. 
Two lines observed at high temperatures are originated from two inequivalent $^{13}$C sites with different $A$ and $K_{\rm orb}$ in a TMTTF molecule (crystallographically inequivalent inner and outer $^{13}$C sites). 
On cooling, each line splits into two lines due to the CO below $T_{\rm CO}\approx$158 K. 
The CO makes two TMTTF molecules on an unit cell inequivalent. 

The amplitude of the charge separation below $T_{\rm CO}$ can be estimated by the measurement of the spin-lattice relaxation rate $1/T_1$. 
In the paramagnetic state, $1/T_1$ is written as $1/T_1=\frac{2\gamma_{\rm n}^2}{\mu_{\rm B}^2}\sum_q(A_qA_{-q})_\perp\frac{{\rm Im}\chi _\perp({\bf q},\omega _{\rm n})}{\omega_{\rm n}}$, where $\gamma_{\rm n}$ is gyromagnetic ratio of the $^{13}$C nuclear, $\mu_{\rm B}$ is Bohr magneton, $A_q$ is the $q$ dependent hyperfine coupling, $\omega_{\rm n}$ is the Larmor frequency ($\sim$86 MHz), and ${\rm Im}\chi _\perp$ is the imaginary part of the dynamical susceptibility perpendicular to the external magnetic field \cite{MoriyaPTP1962}. 
We assume here that the weak site dependence of ${\rm Im}\chi _\perp({\bf q},\omega _{\rm n})$, isotropic $A_{q}$ ($A$), and $A\propto \rho$, where $\rho$ is the charge density, leading that $1/T_1$ is roughly proportional to $\rho^2$.
$A\propto \rho$ is reliable because we consider here just one band ($\pi$ orbital).
The recovery curve of the nuclear magnetic relaxation is obtained for each site by integrating intensity of the spectrum.
As the recovery curve cannot be fitted by a single exponential function near the antiferromagnetic critical region, we used a stretched exponential function $[M(\infty)-M(t)]/M(\infty)=\exp[-(t/T_1)^\lambda]$, where $\lambda$ parametrizes the distribution of $1/T_1$ in order to discuss the temperature dependence of $1/T_1$ in the whole temperature range.  
At 202 K, $1/T_1$ for the higher and lower signals are estimated to be 115 s$^{-1}$ and 45.7 s$^{-1}$, respectively. 
This indicates that the higher signal is from the inner $^{13}$C site in the TMTTF molecule while the lower signal originates from the outer site (see Fig. \ref{fig:CrystalStructure}), since $\rho$ tends to be large near the center of the stacking chain. 
The temperature dependence of $1/T_1$ is shown in Fig. \ref{fig:Freq+T1Ambient}(a). 
$1/T_1$ decreases with decreasing temperature above $T_{\rm CO}$. 
The feature of the temperature dependence around $T_{\rm CO}$ agrees with that of the earlier report \cite{YuJPF2004}. 
We estimated the amplitude of the charge separation by using the ratio of $1/T_1$ for split lines below $T_{\rm CO}$ \cite{COT1}.
In the inset of Fig. \ref{fig:Freq+T1Ambient}(a), the temperature dependence of $\sqrt{(1/T_1^{\rm R})/(1/T_1^{\rm P})}$, where $T_1^{\rm R}$ ($T_1^{\rm P}$) is the relaxation time of rich (poor) site, is shown. 
The ratio of the estimated charge separation is about 3.5 : 1 at 50 K, which is close to the previously obtained ratio of 3 : 1 \cite{YuJPF2004}.

\begin{figure}[b]
\includegraphics[scale=0.35]{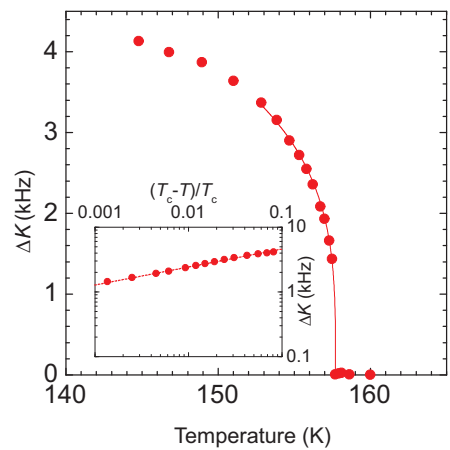}
\caption{Temperature dependence of the line splitting $\Delta K(T)$ of $^{13}$C NMR spectrum of below $T_{\rm CO}$. In the inset, $\Delta K(T)$ is plotted as a function of the reduced temperature. \label{fig:COCritical}}
\end{figure}

Figure \ref{fig:Freq+T1Ambient}(b) shows the temperature dependence of the peak positions of the $^{13}$C NMR spectra; the symbols correspond to those in the spectra in Fig. \ref{fig:SPCAmbient}. 
Near $T_{\rm CO}$ many physical properties in the system are governed by the critical behavior of the CO transition. 
The amplitude of the charge separation is also reflected in the degree of the NMR line splitting $\Delta K(T)$ below $T_{\rm CO}$.
At $T_{\rm CO}$ both $K_{\rm s}$ and $K_{\rm orb}$ for the inner and outer $^{13}$C nuclei will change critically due to the CO, resulting in the line splitting. 
When we assume $K_{\rm s}$ and $K_{\rm orb}$ follow a power law $(T_{\rm CO}-T)^\beta$, where $\beta$ is a symmetry breaking critical exponent, $\Delta K(T)$ is written as  $\Delta K(T)\propto (T_{\rm CO}-T)^\beta$. 
Figure \ref{fig:COCritical} shows the temperature dependence of $\Delta K(T)$ below $T_{\rm CO}$.
The order parameter can be characterized by a critical exponent $\beta$.
$\beta\sim$0.28 is derived by fitting in the temperature range from 149 K to 157.5 K for $T_{\rm CO}$=157.707 K, which is the fitting parameter.
This value is smaller than $\beta$=0.5 for the mean field value and $\beta=0.325\pm0.001$ for the 3D Ising model \cite{LumsdenPRB1998}.
The early works suggested the mean field value \cite{ZamborszkyPRB2002, ChowPRL2000}.
Our precise estimation indicates that $\Delta K(T)$ rises rapidly just below $T_{\rm CO}$, whereas the CO transition seems to occurs almost continuously.
Although, we do not exclude the possibility that the CO transition is of the first order, the rapid rise suggests that there is an another mechanism for the transition, which determines the nature of the phase transition, other than the Coulomb interaction along the one-dimensional chain (e. g. the shift of anions).

On cooling, additional line splitting occurs below $T_{\rm AF}\approx$7.5 K as shown in Fig. \ref{fig:SPCAmbient}(a). 
The symmetric line splitting is attributable to the staggered moment in the AF phase. 
The well-separated peaks indicate that the spin configuration is commensurate with the lattice.
Above $T_{\rm AF}$, four lines originated from the CO merge into two lines with decreasing temperature.
$1/T_1$ for the charge-rich site, $1/T_1^{\rm R}$, and poor site, $1/T_1^{\rm P}$, are separated by fitting the recovery curve of the nuclear relaxation with the double exponential function, $[M(\infty)-M(t)]/M(\infty)=A(\exp[-(t/T_1^{\rm R})]+\exp[-(t/T_1^{\rm P})])$ for two lines. 
$1/T_1$, represented in Fig. \ref{fig:Freq+T1Ambient}, increases below about 30 K and shows a pronounced peak at $T_{\rm AF}$. 
This is caused by the critical slowing down of the antiferromagnetic spin fluctuations, considering $1/T_1$ can detect the correlation of spin fluctuations at Larmor frequency.

\begin{figure}[b]
\includegraphics[scale=0.44]{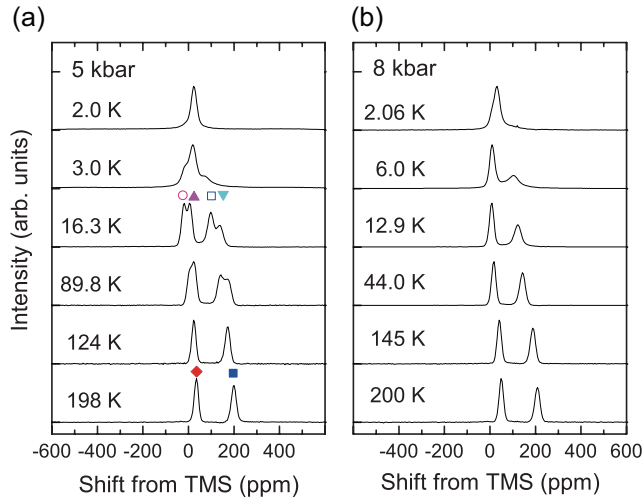}
\caption{(a) Temperature dependence of $^{13}$C NMR spectrum of (TMTTF)$_2$SbF$_6$ at 5 kbar. (b) at 8 kbar. \label{fig:SPC5kbar+8kbar}}
\end{figure}

\subsection{Reduction of $T_{\rm CO}$ under pressure}
\begin{figure}[b]
\includegraphics[scale=0.45]{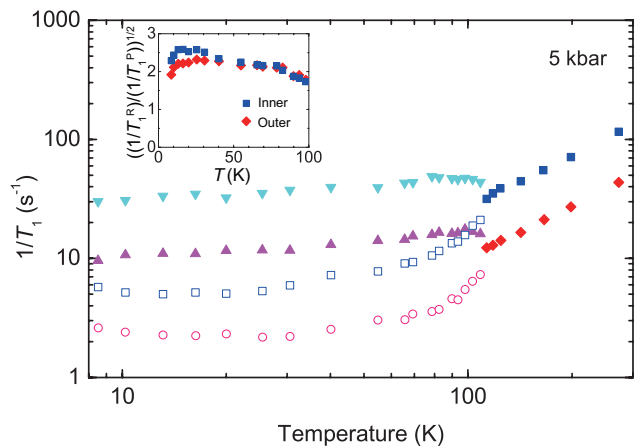}
\caption{Temperature dependence of $1/T_1$ under 5 kbar. \label{fig:T1_5kbar}}
\end{figure}

Next, we discuss the CO transition under pressure. 
Figure \ref{fig:SPC5kbar+8kbar}(a) shows the temperature dependence of the spectrum at 5 kbar. 
At this pressure, the line splitting due to the CO occurs below $T_{\rm CO}\sim$110 K. 
The CO phase transition eventually disappears at 8 kbar as shown in Fig. \ref{fig:SPC5kbar+8kbar}(b). 
This behavior is simply explained by the decrease of $V/t$ by pressure.
The overlapping of the spectra below $T_{\rm CO}$, that suggests the small charge separation, prevents the estimation of the CO order parameter. 
In Fig. \ref{fig:T1_5kbar}, the temperature dependence of $1/T_1$ measured at 5 kbar is shown. 
In the metallic state at high temperatures, the itinerant-electrons mainly contribute to the relaxation. 
In this condition, $1/T_1$ is written as $1/T_1=\pi A^2N(E_{\rm F})^2T$, where $A$ is the average of the $q$-dependent hyperfine coupling constant, and $N(E_{\rm F})$ is the density of the state at Fermi energy $E_{\rm F}$. 
The fact that the value of $1/T_1$ for inner and outer $^{13}$C sites are smaller than that at ambient pressure is attributable to the decrease of $N(E_{\rm F})$ due to the increase of the band width $W (\propto t)$. 
Below $T_{\rm CO}$, we estimate the charge separation by the ratio of $1/T_1$ in the same manner with the ambient pressure (see the inset of Fig. \ref{fig:T1_5kbar}). 
$\sqrt{(1/T_1^{\rm R})/(1/T_1^{\rm P})}$ increases below $T_{\rm CO}$, and saturates near 10 K. 
The charge separation is roughly in the ratio of 2 : 1, which is lower than the value at ambient pressure. 
This also indicates that the CO is destabilized by the pressure.

\begin{figure}[b]
\includegraphics[scale=0.5]{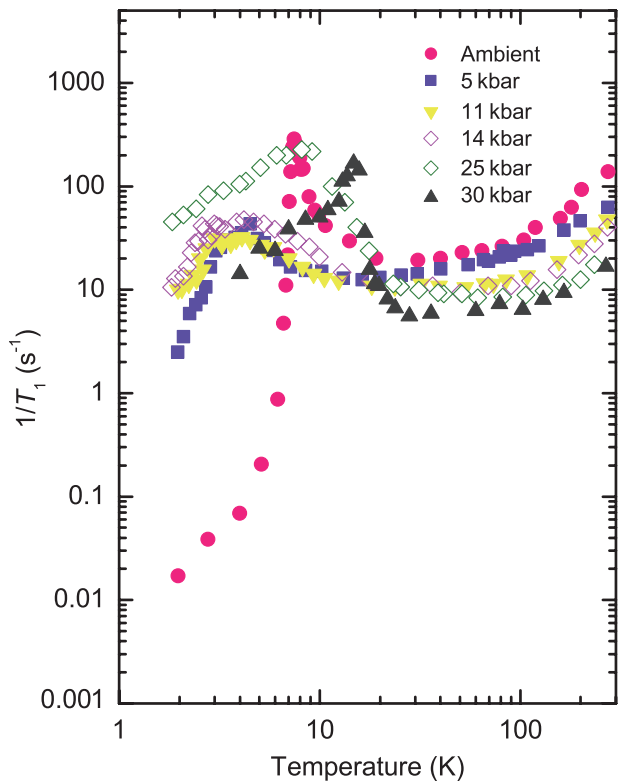}
\caption{Temperature dependence of the spin-lattice relaxation rate, $1/T_1$, at six different pressures. \label{fig:T1Pressdep}}
\end{figure}

\begin{figure}[b]
\includegraphics[scale=0.44]{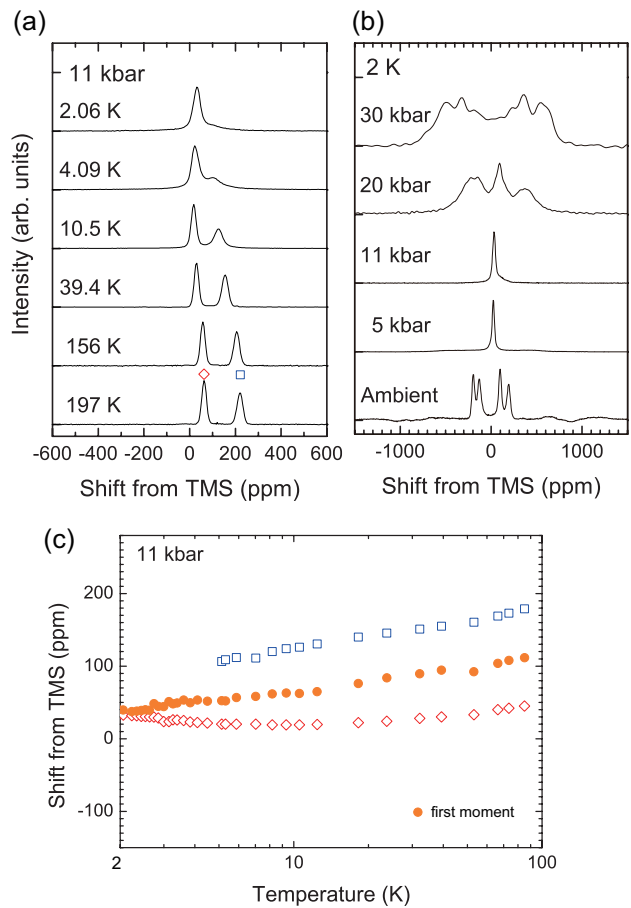}
\caption{Temperature dependence of the $^{13}$C NMR spectrum of (TMTTF)$_2$SbF$_6$ under (a) 8 kbar and (b) 11 kbar. (c) Temperature dependences of the peak position and center of gravity (first moment) of the $^{13}$C NMR spectrum at 11 kbar. The symbols are the same as those in the spectrum in (a). \label{fig:SPC11kbarPressureFreq11kbar}}
\end{figure}

\subsection{Spin-gap transition under pressure}
The spectrum obtained at 5 kbar exhibits completely different feature from that at ambient pressure.
Below about 3 K, the spectrum forms a sharp peak. 
The diminishing of the site dependence of the NMR shift is caused by the reduction of spin susceptibility, $\chi_s\to 0$, considering $K_{\rm s}=A\chi_s$. 
AsF$_6$, PF$_6$, and I salts have similar spectral shape below $T_{\rm sP}$ (or spin-singlet transition temperature $T_{\rm SS}$) \cite{BrownPRL1998, BrownSM1999, ZamborszkyPRB2002, FujiyamaJPSJ2006, NakamuraJPSJ2007, FurukawaPRB2011}. 
A previous $^{13}$C 2D NMR study obtained split spectra under 6 kbar on the dipolar coupling axis, that is attributed to the sP state \cite{YuPRB2004}. 
The single line spectrum is also observed at 8 kbar and 11 kbar as shown in Fig. \ref{fig:SPC5kbar+8kbar}(b) and \ref{fig:SPC11kbarPressureFreq11kbar}(a). 
The temperature dependence of $1/T_1$ under various pressures is shown in Fig. \ref{fig:T1Pressdep}. 
$1/T_1$ was defined as the initial slope of the relaxation curve of nuclear magnetization after saturation.
This estimation characterizes the volume average of $1/T_1$.
The striking feature is the suppression of the critical increase in $1/T_1$ observed near $T_{\rm AF}$. 
$1/T_1$ shows a drop below about 3--4 K without significant peak structure, which is attributable to a spin-gap transition.
Although it is difficult to conclude that the spin-gap is opened or not, because the temperature range measured below the anomaly is too narrow that prevents the characterization of the gap structure, we here attribute this behavior to a spin-gap transition considering the similarity with that in the other TMTTF complexes.
The gap size is $\Delta/k_{\rm B}\sim 8$ K, estimated by using the data of $1/T_1$ under 11 kbar just below the spin-gap transition temperature and assuming that $1/T_1$ follows an activation type $1/T_1\propto \exp(-\Delta/k_{\rm B}T)$.

However, the NMR shift has different character.
Figure \ref{fig:SPC11kbarPressureFreq11kbar}(c) shows the temperature dependence of the resonance frequency at the peak position and the center of gravity of the spectra (first moment) at 11 kbar. 
The first moment monotonically decreases with decreasing temperature.
In TMTTF salts, as the Knight shift is negative in the paramagnetic state, a positive shift is expected below sP transition \cite{FujiyamaJPSJ2006, NakamuraJPSJ2006, NakamuraJPSJ2007, IwasePRB2010}. 
The reason for the shift to lower frequencies, which is the opposite direction for the nonmagnetic spin-singlet state, is not clear at this moment. 
Our observation indicates that there is no reduction in $\chi_{\rm s}$.
Even though the local spin susceptibility at the outer $^{13}$C site (indicated by the diamond symbols in Fig. \ref{fig:SPC11kbarPressureFreq11kbar}(c)) appears to decrease with decreasing temperature below 10 K, the first-moment shows no anomaly at the temperature. 
It is possible that $K_{\rm orb}$ (or Knight shift origin) is modulated by the external pressure. 
As the result, $K_{\rm orb}$ becomes lower and the sign of $A$ changes from negative to positive. 
The shift origin can be estimated by an analysis of the $K_{\rm obs}$ versus spin susceptibility $\chi$ plot under pressures.
The investigation of the spin susceptibility $\chi_{\rm s}$ under pressure is under way.




The coexistence of the CO and sP orderings has been claimed evidenced by the broadened spectrum of $^{13}$C NMR in (TMTTF)$_2$PF$_6$ at 21.0 T \cite{ZamborszkyPRB2002}. 
Our experiment could not detect the coexistence as the magnetic field in our experiment is 9 T which is lower than the critical field $B_{\rm c}$ for the triplet excitations. 


\begin{figure}[t]
\includegraphics[scale=0.48]{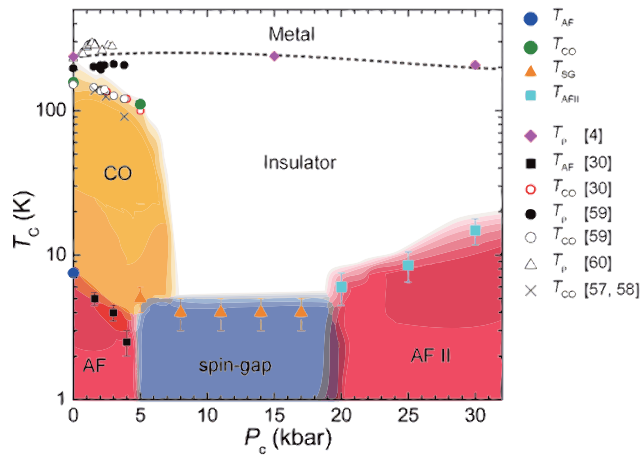}
\caption{Pressure--temperature phase diagram of (TMTTF)$_2$SbF$_6$ constructed based on NMR measurements at hydrostatic pressures. Data from refs. \cite{YuPRB2004, NagasawaJPF2004, NagasawaSSC2005, NagasawaJPSJ2008, ItoiJPSJ2008, ItoiPC2010} are also shown. \label{fig:PhaseDiagram}}
\end{figure}

\subsection{Antiferromagnetic ordering under pressure}
Above 20 kbar, the spectrum shows a different feature at low temperatures. 
Figure \ref{fig:SPC11kbarPressureFreq11kbar}(b) shows the pressure dependence of the $^{13}$C NMR spectrum at $2.0\pm 0.1$ K. 
The spectrum exhibits significant broadening above 20 kbar. 
This behavior indicates that the system enters into the expected AFII phase. 
$1/T_1$, estimated at the center of the spectrum, shows the critical increase near the transition temperature $T_{\rm AFII}$ as shown in Fig. \ref{fig:T1Pressdep}, supporting the AFII transition. 
The spectral shape is different from that for AFI phase, in which four peaks are well separated. 
The spectral profile exhibits some characteristics of weak incommensurability with the lattice, namely a wide distribution ($\pm$500 ppm) at 30 kbar, a substantial spectral density at the center of the spectrum, and a narrow tail compared to that at 20 kbar. 
$1/T_1$ decreases below $T_{\rm AFII}$ gradually compared to the reduction at ambient pressure. 
If the ground state is incommensurate-spin density wave (IC-SDW), the {\it gradual} decrease of  $1/T_1$ below $T_{\rm AFII}$ is attributable to the phason mode, which dominates spin-lattice relaxation in the IC-SDW phase \cite{BarthelEL1993}. 
$1/T_1$ of the side wings of the spectrum for IC-SDW phase corresponds to the relaxation due to the amplitude modes.
As we estimated $1/T_1$ only at the central part of the spectrum, we cannot discuss the difference of the $1/T_1$ between the central part and the wing part of the spectrum at present. 
This analysis will be the future plan.
The one-dimensional IC-SDW phase is expected to have a sinusoidal NMR spectrum, whereas the observed line shape at 30 kbar is not sinusoidal, suggesting that commensurate domains remain. 
In fact, the ground state is expected to be the commensurate AF phase driven by the spin interaction because the charge gap is opened at 30 kbar ($T_{\rho}>0$) \cite{ItoiJPSJ2008} for which the spin-localized picture is appropriate. 
Our observation suggests that the system is near the insulator--metal boundary above $T_{\rm AFII}$.
The AF ordering patterns of AF and AFII in (TMTTF)$_2$SbF$_6$ is an important topic for future studies.

\subsection{Phase diagram}
In (TMTTF)$_2$SbF$_6$, the behaviors of the AF and CO transitions under low pressures below 5 kbar have been established \cite{YuPRB2004}.
Above the $T_{\rm CO}$, a metal-insulator crossover has been reported \cite{NagasawaJPF2004, NagasawaSSC2005, NagasawaJPSJ2008, ItoiJPSJ2008, ItoiPC2010}.
Figure \ref{fig:PhaseDiagram} shows a phase diagram of (TMTTF)$_2$SbF$_6$ that summarizes the results of the present study. 
$T_{\rm CO}$ decreases with increasing pressure, that is consistent with the previous study \cite{YuPRB2004}.
At 8 kbar, we could not find the CO transition.
Under 5 kbar $<P<$ 20 kbar, the system shows the SG phase transition below about 4 K.
The existence of the singlet phase has been suggested in the ref. \cite{YuPRB2004}, though the transition temperature has not been established.
If this state is a non-magnetic sP phase, the sequence of the ground states contradicts with the pressure effect on the dimensionality of the system \cite{ItoiJPSJ2008, FurukawaPRB2011}.
In our experiment, the expected reduction of the Knight shift has not been observed below $T_{\rm SG}$.
As pointed in ref. \cite{FurukawaPRB2011}, the interchain coupling probably plays an important role for the realization of the unconventional spin-gap phase. 
We observed the evidence of the AFII phase transition, which has been expected in ref. \cite{YuPRB2004}.
The phase transition temperature of AFII phase, $T_{\rm AFII}$, is defined as the peak of $1/T_1$. 
$T_{\rm AFII}$ increases with increasing pressure as $dT_{\rm AFII}/dP\sim +8.5$ K/kbar. 
In the charge-gap state, the increase in $T_{\rm AF}$ can be attributed to an increase in interchain spin coupling with pressure \cite{KlemmePRL1995} as discussed in (TMTTF)$_2$Br at moderate pressures of below 5 kbar \cite{KlemmePRL1995, HisanoSM1999}. 
The increase of the dynamic susceptibility $\chi(2k_{\rm F})$ near $T_{\rm AFII}$ becomes significant with increasing pressure as seen in the $1/T_1$ data ($1/T_1\propto {\rm Im}\chi (2k_{\rm F})$ near $T_{\rm AFII}$).
On further increasing the pressure a reduction in $T_{\rm AFII}$ and a clear IC spin density wave (SDW) phase are expected to be observed due to imperfect nesting in the SDW state \cite{FuseyaJPSJ2007}. 
In this situation $\chi (2k_{\rm F})$ will reduce with pressure on approaching the SC phase as observed in (TMTSF)$_2$PF$_6$ \cite{CreuzetSM1987}.
The observation of the widely distributed superconducting phase (5.4 $<P<$ 9 GPa) \cite{ItoiJPSJ2008} is consistent with this scenario.

\section{Conclusion}
In conclusion, we conducted $^{13}$C NMR under pressures to study the magnetic properties of quasi-one-dimensional complex (TMTTF)$_2$SbF$_6$. 
NMR measurements revealed the magnetic property and the criticality of the CO transition. 
For 5 $< P < $20 kbar, the observations of the rapid decrease of $1/T_1$ and the spectral change suggest that the ground state is the SG state.
We found an AFII phase above 20 kbar.  
To our knowledge this is the first report of the observation of the AFII phase in (TMTTF)$_2$SbF$_6$. 
$T_{\rm AFII}$ increases with increasing pressure in the measured pressure range. 
The broadened spectrum above 30 kbar suggests that the magnetic moment tends to be incommensurate with the lattice at higher pressures.

\section*{Acknowledgments}
The authors wish to thank Y. Shimizu, T. Ito, H. Seo, M. Itoi, K. Yonemitsu, P. Monceau, T. Takahashi, and H. Fukuyama for valuable discussions. This study was supported by Grants-in-Aid for Scientific Research (B) (No. 20340095) from JSPS, Scientific Research on Innovative Areas (No. 21110523) from the Ministry of Education, Culture, Sports, Science and Technology, Japan. F. I. was partially supported by Grants-in-Aid for Scientific Research for Young Scientist (B) (No. 23740273). K. F. was partially supported by Grants-in-Aid for Scientific Research \lq`New Frontier of Materials Science Opened by Molecular Degrees of Freedom\rq' (No. 21110523) and for Young Scientists (A) (No. 21685021) from JSPS.

\end{document}